\documentclass{optica-article}

\journal{opticajournal} 

\usepackage{graphicx}
\usepackage{dcolumn}
\usepackage{bm}
\usepackage[utf8]{inputenc}
\usepackage[T1]{fontenc}
\usepackage{float}
\usepackage{amsfonts}
\usepackage{multirow}
\usepackage{booktabs}
\usepackage{makecell}
\usepackage{tabularx}
\usepackage{setspace}
\usepackage{url}
\usepackage{amsmath}

\usepackage{amssymb}

\articletype{Research Article}

\usepackage{lineno}

\begin{document}

\title{Precision spectroscopy and frequency determination of the hyperfine components of the P(63)~4-4 transition of molecular iodine near 652~nm}

\author{Shamaila Manzoor,\authormark{1,3} Mauro Chiarotti,\authormark{1,3} Samuel A. Meek,\authormark{5} Gabriele Santambrogio,\authormark{2,3,4} and Nicola Poli\authormark{1,3,*},}

\address{\authormark{1}Dipartimento di Fisica e Astronomia, Universit\`a degli Studi di Firenze, Via G. Sansone 1, 50019 - Sesto Fiorentino, Italy\\
\authormark{2}Istituto Nazionale di Ricerca Metrologica, INRIM, Via Nello Carrara, 1
 50019 Sesto Fiorentino, Italy\\
\authormark{3}Laboratorio Europeo di Spettroscopia Non-Lineare, LENS  Via N. Carrara 1, I-50019 Sesto Fiorentino, Italy\\
\authormark{4}CNR-INO, Istituto Nazionale di Ottica, Via N. Carrara 1, I-50019 Sesto Fiorentino, Italy\\
\authormark{5}Homer L. Dodge Department of Physics and Astronomy, Center for Quantum Research and Technology, University of Oklahoma, 440 W. Brooks St., Norman, OK 73019, USA}

\email{\authormark{*}nicola.poli@unifi.it} 


\begin{abstract*} 
We report the observation of the hyperfine spectrum of the weak P(63)~4-4 line of the $B-X$ electronic transition of molecular iodine $^{127}$I$_2$ near 652.4~nm, using frequency-modulated saturated absorption spectroscopy. Through the precise measurements of the absolute frequencies of hyperfine components, we estimate electric quadrupole and magnetic spin-rotation constants. Additionally, we determine the center of gravity of the P(63) transition of the 4~-~4 vibrational band, resulting in a 250-fold improvement in the precision of its position. We also note an interesting overlap of the hyperfine transitions of P63(4-4) line with the UV $^1$S$_0$~-~$^3$P$_1$ narrow intercombination transition of cadmium atoms, which occurs near the second harmonic of the master laser radiation, corresponding to 326.2~nm. This study contributes to updating the iodine atlas, improving the precision of the empirical formulae, and providing an important frequency reference for precision spectroscopy of the narrow intercombination transition of atomic cadmium.

\end{abstract*}


\section{Introduction}
The absorption spectrum of molecular iodine (I$_2$), spans from 7220~cm\textsuperscript{-1} to 23800~cm\textsuperscript{-1} (equivalent to 420.168~nm to 1385.042~nm)~\cite{Luc85}. Significant research is centered on the transitions within the (B$ ^{3}\Pi_{0u}^{+} \longleftarrow X^{1}\Sigma _{g}^{+}$) manifold due to its broad spectral coverage spanning from green to near-infrared. Although the I$_2$ molecule serves as a useful frequency reference, its Doppler-broadened absorption lines exhibit asymmetry due to hyperfine structures and potential overlapping transitions, leading to temperature-dependent lineshapes. These Doppler-broadened lines have a few GHz linewidths, limiting their accuracy as frequency references for lasers stabilized to their relative centers of gravity, making them unsuitable for direct use as precise frequency standards. 

However, the Doppler-broadened lines can be resolved using methods like saturated absorption spectroscopy, Doppler-free frequency-modulation spectroscopy, or supersonic molecular beam technique~\cite{demtröder2008}. The hyperfine transitions of these Doppler lines have line widths of the order of MHz.
 As a result, several groups have conducted precise frequency measurements of the hyperfine structures of the selected iodine lines across a range of rovibrational levels for calibration purposes or laser stabilization. This has been demonstrated over a wide range of wavelengths from around 500~nm to 830~nm~\cite{Cheng19, Hong09, Xie19, Liao2010, Wu2022, Schuldt2017, Sakagami2020, Ikeda2022, Matsunaga2024}. Such hyperfine transitions can be conveniently used as frequency references and have been exploited, for instance, for experiments on thallium atoms at 535~nm~\cite{Shie2013}, helium atoms at 541 nm~\cite{Pastor2000}, ytterbium atoms at 556~nm~\cite{Tanabe2022}, lithium atoms at 647~nm~\cite{Huang2018}, barium ions at 650~nm~\cite{Xie19}, calcium atoms at 657~nm~\cite{Morinaga89}, iron atoms at 716~nm~\cite{Huet2013}, and francium atoms at 718~nm~\cite{Dube04}. Some of the hyperfine components of these iodine transitions have been endorsed by the Comité International des Poids et Mesures (CIPM) as frequency standards for the practical implementation of meter, the SI unit of length ~\cite{Quinn2003}. Iodine-stabilized lasers are also used as reference frequency markers for optical frequency combs and also as a flywheel in the frequency measurements~\cite{YeJun2001, Hong2005}.

Beyond their important role as frequency standards in the optical domain, precise spectroscopic investigation of these transitions provides invaluable information regarding nuclear properties and atomic structures~\cite{Campbell2016}. To interpret the molecular spectra, we use an effective Hamiltonian model that accounts for hyperfine interactions coupling the various rotational and spin states; the parameters for this model follow those described by Broyer et al.~\cite{Broyer78}. The Hamiltonian includes the contributions of four different interaction terms, which are proportional to four independent constants: the electric quadrupole coupling constant~$eQq$, the spin-rotation~$C$,  the tensorial spin-spin~$d$, and the scalar spin-spin~$\delta$ parameter, respectively. The determination of these hyperfine constants across different rovibrational transitions contributes to the development of the empirical interpolation formulae~\cite{Bodermann2002}. These formulae describe the overall molecular iodine spectrum across various rotational and vibrational levels within the $B-X$ electronic state. The predicted uncertainty of these empirical formulae varies across different rovibronic molecular transitions, primarily depending on the availability of the high-precision measurements for those specific transitions. Particularly, the availability of precise experimental data for hot bands is sparse\cite{Edwards1996, Morinaga89} due to the weak nature of these transitions. Furthermore, by updating these empirical interpolation formulae with high-precision hyperfine constants across different rovibrational levels, it becomes possible to accurately determine the positions of the hyperfine components of the spectrum across the $B-X$ manifold, even if they have not yet been observed. 

This paper presents the first observation and precise frequency determination of the hyperfine transitions of the weak line P(63) of the 4~-~4 band centered at 652.4~nm of the iodine molecule. Interestingly, the second harmonic of the light used for spectroscopy of this particular I$_2$ line overlaps with the  $^1$S$_0$ - $^3$P$_1$ narrow intercombination transition of cadmium (Cd) in the UV, precisely at 326.2~nm. This overlap suggests that iodine could serve as an effective optical frequency reference for frequency-doubled UV laser systems in this particular region, avoiding cumbersome UV spectroscopy setups and the resulting loss in UV power. Specifically, the precise frequency determination of the hyperfine structures of the rovibrational transition of iodine presents significant potential for enhancing precision spectroscopy experiments on Cd, which is essential for testing theories beyond the Standard Model~\cite{Ohayon2022}. This method is particularly advantageous for setups without access to optical frequency combs, providing a simplified method for optical frequency calibration. In the future, this transition may also facilitate the development of second-stage laser cooling of Cd using its narrow intercombination transition at 326.2~nm~\cite{Bandarupally23, Gibble2024}. Indeed, Cd atoms have recently gained significant attention due to their eight stable isotopes, allowing for numerous interesting experiments to be performed, including precision tests of fundamental physics and atom interferometry~\cite{Tinsley22}. Throughout the paper we have used the open-source software PGOPHER~\cite{Western17} for simulating the rovibrational and the hyperfine transitions. PGOPHER is a versatile software that can simulate and fit the rotational, vibrational, and electronic spectra of the various molecular types, including linear, symmetric top, and asymmetric top molecules. The iodine molecule, being linear, can be analyzed using this software as it employs a model containing built-in $\hat{N}^{2}$ Hamiltonian designed for diatomic molecules~\cite{Brown79}. As a visual tool, it enables detailed investigation of hyperfine structures through its graphical representations of fits and residuals, which greatly simplifies data interpretation. It accounts for both Gaussian and Lorentzian contributions to the linewidth. Beyond estimating the hyperfine parameters $eQq$, $C$, $d$ and $\delta$, frequencies of the centre of gravity of the Doppler broadened transitions can also be estimated. Further details regarding the basis set, quantum numbers, and Hamiltonian model for linear molecules, and the spectral fitting process in PGOPHER can be found in Ref.~\cite{Western17}. 

The structure of the paper proceeds as follows: Section~\ref{sec:theo} presents the theoretical model and the expected spectrum of the hyperfine transitions. Section~\ref{sec:exp} details the experimental setup. Section~\ref{sec:expres} presents the experimental spectra, the absolute frequency measurements of the splittings, and hyperfine constants from two different fitting procedures, accompanied by discussions of the investigated lines. Finally, the conclusions drawn from the studies are summarized in the last section.


\section{\label{sec:theo}Theoretical Background}
\begin{figure}[t]
\centering
\includegraphics[width=0.9\textwidth]{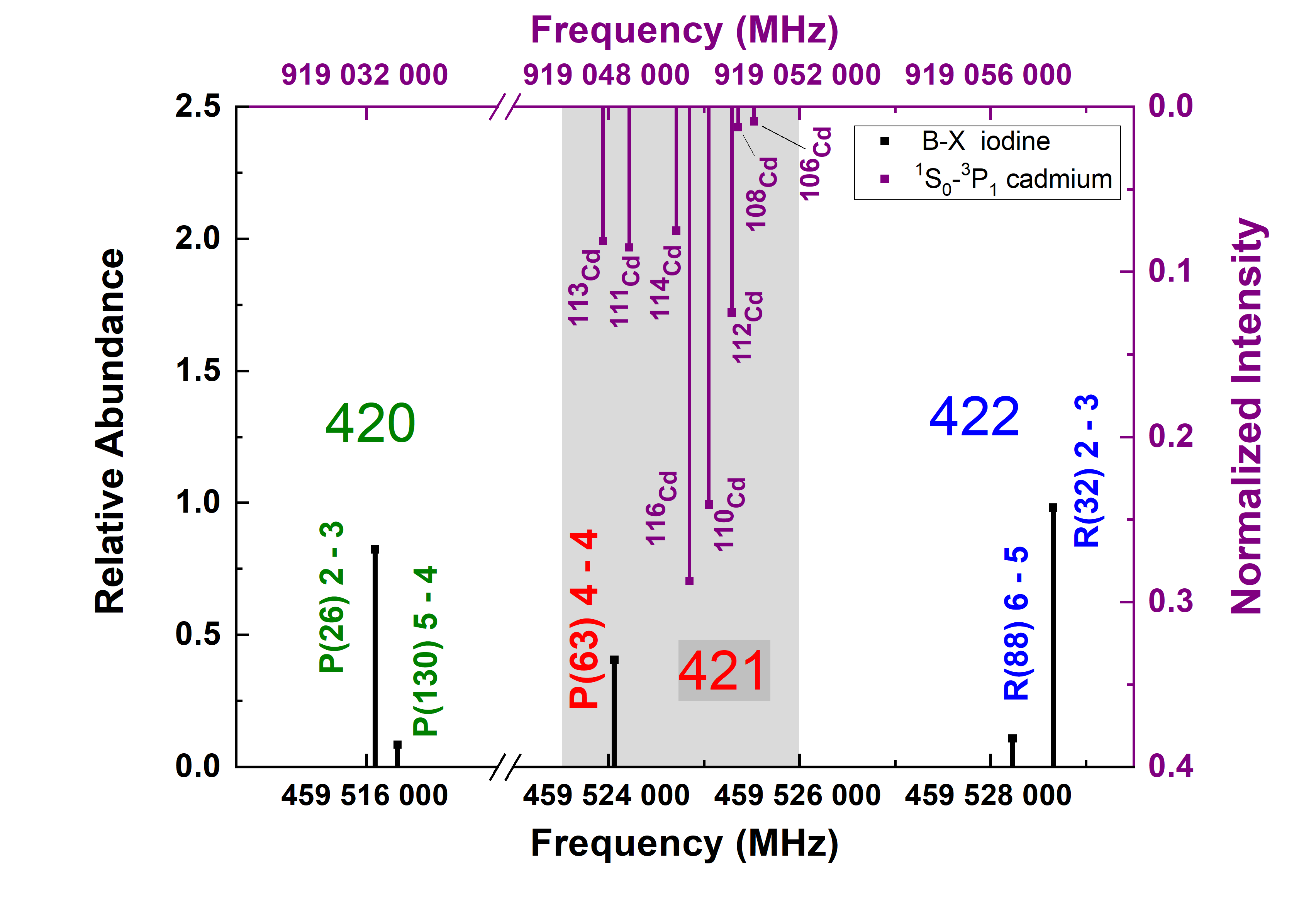}
\caption{Simulated rovibrational spectrum of the $B-X$~I$_2$ transition around 652.397~nm~(approximately 459~524~600~MHz). The second harmonic of this spectral region overlaps with the Cd $^1$S$_0$~-~$^3$P$_1$ intercombination transition at 326.2 nm. The top x-axis shows the transition frequencies (\(2\nu\)) of the eight stable isotopes of Cd~\cite{Manzoor22}. The vertical black droplines indicate the center of gravity of the iodine lines simulated at T~=~293~K using rovibrational constants from the I$_2$ atlas. Labels 420, 421, and 422 correspond to atlas lines representing these transitions~\cite{Luc85}. The grey-shaded area represents the portion of the spectrum expanded in Figure~\ref{fig:der}}.
\label{fig:simspectra}
\end{figure}

With rovibronic constants from Ref~\cite{Luc85}, we use PGOPHER to calculate the spectrum of the iodine molecule around 652.397~nm~(approximately 459~524~600~MHz) at a temperature of 293 K. This spectrum is shown in Figure~\ref{fig:simspectra}. The vertical black droplines represent the centers of gravity of the calculated transition frequencies for the atlas lines 420, 421, and 422. Lines~420 and 422 contain four and five rovibrational transitions, respectively, though only two transitions for each line are shown in the figure. Line~421 is identified as  B $ ^{3}\Pi_{0u}^{+} \longleftarrow X ^{1}\Sigma _{g}^{+}$~P(63)~4-4 transition. Despite its low intensity, this transition is notable for several reasons: it is easily identifiable, being separated by approximately 7.8~GHz from the nearby P(130)~5-4 transition and 4.2~GHz from the R(88)~6-5 transition. Additionally, the second harmonic of this transition is very close to the Cd $^1$S$_0$ - $^3$P$_1$ intercombination transition at 326.2~nm, graphically shown in Figure~\ref{fig:simspectra}.

The center of gravity of the Doppler-broadened line depends on the relative intensities of all hyperfine components. Therefore, only limited accuracy can be obtained from Doppler-broadened measurements. Instead, we measure the absolute frequency of the hyperfine structures directly. This is because the center of the gravity of a Doppler-broadened line, where hyperfine components are blended, may not accurately represent the true center of gravity of the measured hyperfine structures. Measuring the absolute frequency of the hyperfine structures is essential to enhance the accuracy of the iodine atlas, thereby making it a reliable standard. This provides an additional motivation for the extensive study presented here.

For the evaluation of the hyperfine constants of the observed transitions, we use the Hamiltonian for the hyperfine interactions, written as:
\begin{align}
\label{eq:H}
 \hat{H}_{\text{eff}} & = \hat{H}_{HFS} + \hat{H}_{\textrm{R}} 
\end{align}
$\hat{H}_{HFS}$ is given by:
\begin{align}
\label{eq:Ht}
\hat{H}_{HFS} = eQq \sum_{\alpha = 1}^2 \frac{\sqrt{6} T^2_{q=0}(\hat{I}_\alpha,\hat{I}_\alpha)}{4 I_\alpha (2 I_\alpha - 1)} + C \hat{J}\cdot\hat{I}
 - d \sqrt{6} T^2_{q=0}(\hat{I}_1,\hat{I}_2) + \delta \hat{I}_1\cdot\hat{I}_2
\end{align}

where, $eQq$, $C$, $d$, and $\delta$ are the (effective) nuclear electric quadrupole, spin-rotation, tensorial spin-spin, and scalar spin-spin hyperfine constants, respectively. The eigenvalues can be calculated by diagonalizing the Hamiltonians. In calculating the electric quadrupole and tensorial spin-spin interactions, non-diagonal contributions are taken into account, which involve changes in nuclear spin $\Delta I$ and changes in molecular rotation $\Delta J$ in the Hamiltonian matrix. Therefore, it is necessary to add the rotational contributions before diagonalizing the Hamiltonians given by:

\begin{align}
\hat{H}_{\textrm R} = T + B \hat{J}^2 - D \hat{J}^4 + H \hat{J}^6 + L \hat{J}^8 + M \hat{J}^{10}
\end{align}
Here, $T$ represents the band origin, and $B$, $D$, $H$, $L$, $M$ are the rotational constants, taken from the literature~\cite{Luc85}. 

\begin{figure*}[t]
\centering
\includegraphics[width=\textwidth]{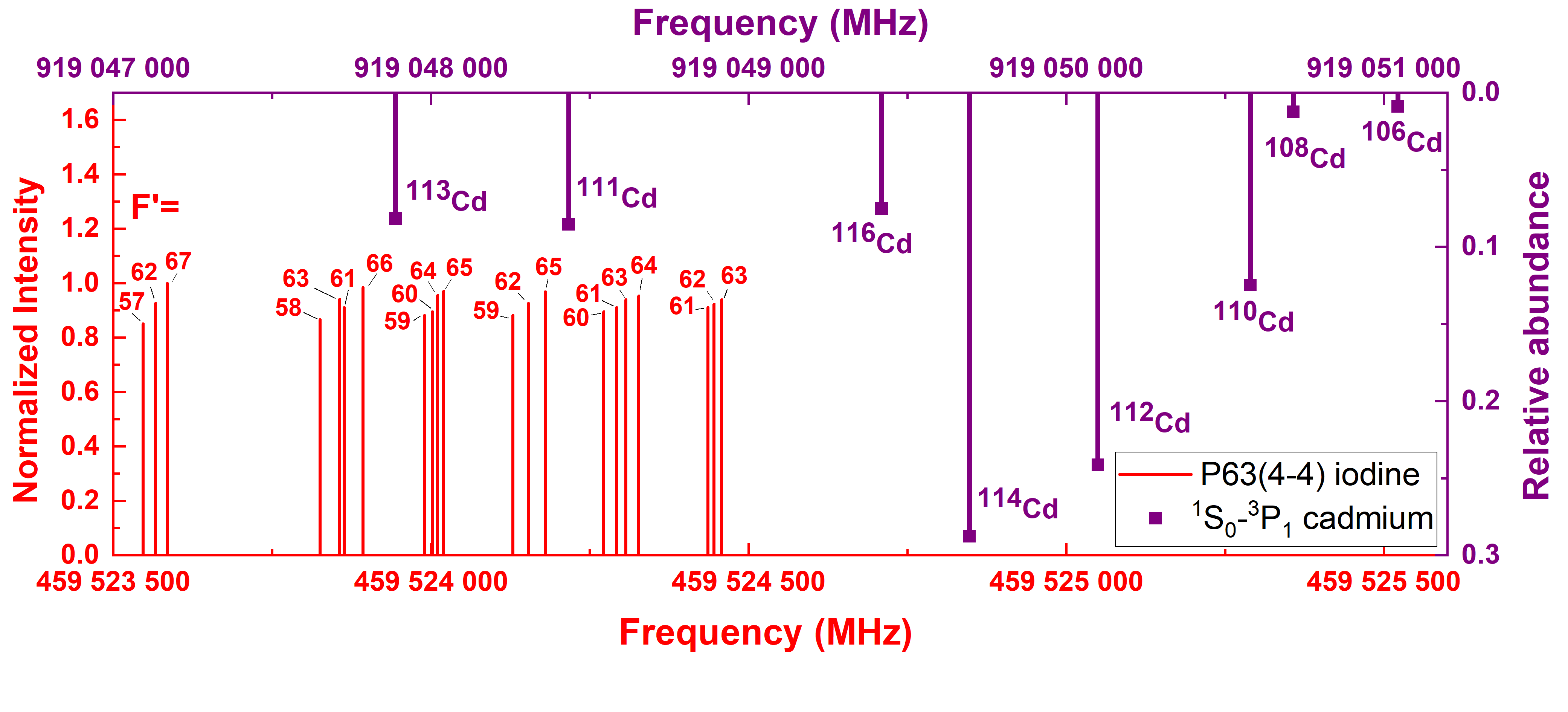}
\caption{Calculated absolute frequencies of the hyperfine components of the P(63)~4-4 transition of I$_2$ around 652.397 nm~(approximately 459~524~600~MHz), identifying all 21 expected components. The hyperfine components are simulated using PGOPHER software~\cite{Western17}. The hyperfine constants were set to $\Delta eQq=1957.67$~MHz and $\Delta C=0.0214$~MHz~\cite{Bodermann2002} and temperature $T_{rot}=293$~K. The rovibronic constants were taken from Ref~\cite{Luc85}. The plot also shows the overlap with the second harmonic of the Cd intercombination transition at 326.2~nm. The top x-axis shows the transition frequencies (\(2\nu\)) of the eight stable isotopes of Cd~\cite{Manzoor22}. The plot is zoomed-in view of the grey area of Figure~\ref{fig:simspectra}.}
\label{fig:der}
\end{figure*} 

The significant energy separation between two adjacent vibrational levels with different $J$ values ensures that the hyperfine coupling between distinct vibrational levels within the B state is negligible, even if they exhibit a non-zero Franck–Condon overlap. In rovibronic transitions with large $J$ values, the primary selection rules are $\Delta F=\Delta J$~(only main lines). Thus, we expect to measure primarily the spectral lines corresponding to $\Delta F = \Delta J$. Specifically, considering that total angular momentum quantum number $J$ in the ground state $X$ is odd and in the excited state $B$ is even, with ortho-nuclear spin states \(I = 1, 3, 5\), we expect the presence of 21 hyperfine components that arise due to the combinations of $J$ and $I$.

A simulation of the expected hyperfine transitions is presented in Figure~\ref{fig:der}. The simulation was performed using the software PGOPHER~\cite{Western17}, employing rotational constants and origin value of the 4~-~4 vibrational band from Ref.~\cite{Luc85}, along with the nuclear quadrupole coupling constants \(eQq'\), \( eQq'' \) and spin-rotation constant \(C'\) derived from the empirical formula in Ref.~\cite{Bodermann2002} with \(C''\)
set to zero. Throughout this paper, single primes denote the upper state, while double primes denote the ground state. The rotational temperature is set at room temperature $T_{rot}~=~293$~K. The hyperfine transitions (21 in total) are constituted by 6 sets of triplets and quadruplets, spanning approximately 910~MHz. According to the constants provided in the existing literature~\cite{Luc85}, the center of gravity of the P(63)~4-4 transition is expected to be approximately 459~524~056~MHz. 
 
The experimental data was treated using PGOPHER. The obtained results, along with the full set of estimated constants, are presented in Section~\ref{sec:expres}.


\section{\label{sec:exp}Experimental Setup}

\begin{figure}[t]
\centering
\includegraphics[width=0.95\textwidth]{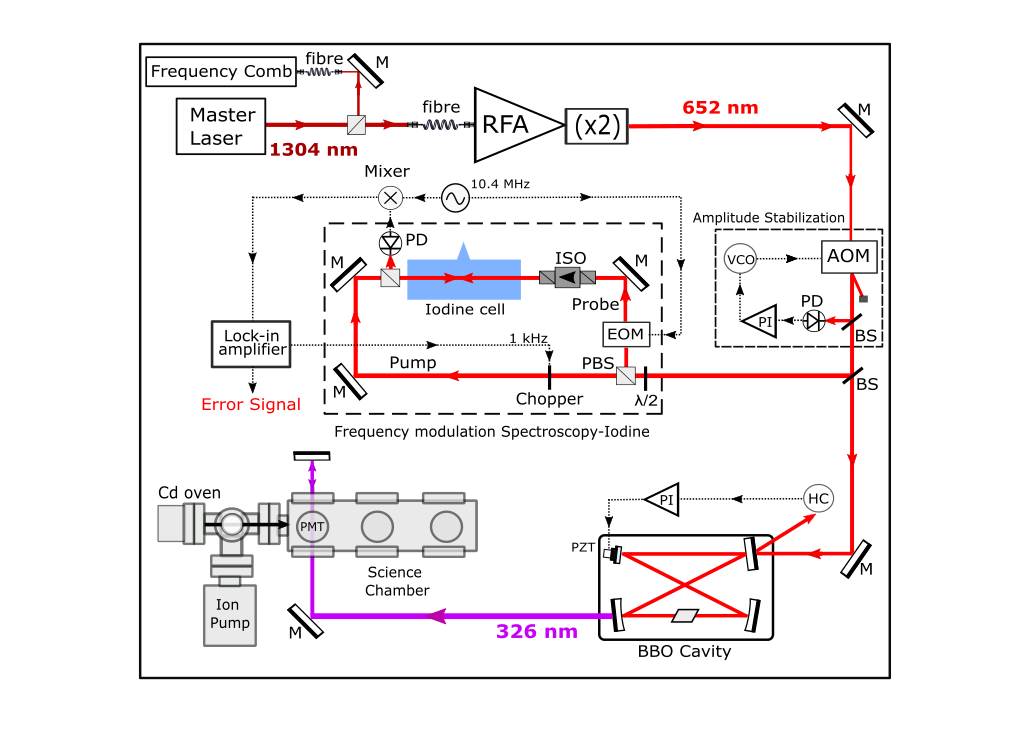}
\caption{Schematics of the experimental setup. M - mirrors, ISO - optical isolator, $\lambda/2$ - Half Wave Plate, PBS - Polarizing Beam Splitter, AOM - Acousto-Optical Modulator, EOM - Electro-Optical Modulator, PD - Photodiodes, HC - Hansch-Couillaud Locking, BBO - Beta-Barium Borate non-linear crystal.}
\label{fig:setup}
\end{figure}

A sketch of the experimental setup is presented in Figure~\ref{fig:setup}. The setup is mainly composed of a laser system, delivering continuous-wave frequency-stabilized laser radiation respectively at 1304.8~nm, at the second (652.4~nm), and the fourth harmonic of this radiation (326.2~nm). The setup is completed by an I$_2$ spectroscopy cell and a Cd atomic beam. Finally, a self-referenced infrared optical comb stabilized by a quartz oscillator slaved to the global positioning system (GPS) is used for all the optical frequency measurements presented in the rest of the paper. Part of the frequency-stabilized light from the master source at 1304.8~nm is sent directly to the comb using a 200~m-long fiber link.

The laser system used in this experiment has been described previously~\cite{Manzoor22}. In brief, the master laser is a home-built tunable laser operating at 1304.8~nm. For frequency stabilization of the master laser, 1~mW of the beam is picked up and sent to the medium finesse ($F\sim10^4$) Fabry-Perot cavity using the Pound-Drever Hall locking scheme. Around 50~mW of the power is sent to a Visible Raman Fibre Amplifier~(VRFA) through a polarization-maintaining fiber. The output of the VRFA laser (up to 4~W at 652.4~nm) is sent through an acousto-optical modulator (AOM) for amplitude stabilization and separated into two beams, the first one for direct spectroscopy on the I$_2$ molecule, the second towards the successive frequency doubling stage to produce the UV light required for spectroscopy on Cd atoms.

As shown in Figure~\ref{fig:setup}, to perform frequency-modulated saturated absorption spectroscopy, two counter-propagating beams are produced, pump and probe, with power of 10~mW and 3~mW, diameter at the center of the 10-cm-long I$_2$ cell of 0.74~mm and 0.77~mm, respectively.  I$_2$ cell. The phase of the probe beam is modulated by an electro-optical modulator~(EOM) adding sidebands separated by 10.4~MHz from the optical carrier. The phase-modulated probe beam then passes through a 30~dB optical isolator and enters the iodine cell, while propagating through the cell in the opposite direction with polarization orthogonal to the probe beam. We find the combination of an additional polarization beam splitter and the optical isolator useful to avoid detrimental optical interference from the counter-propagating intense pump beam on the EOM. 

A fast photodetector is used to detect the AC beat signal at the modulation frequency on the probe beam. The beat signal arises closer to the Doppler-free absorption line due to the saturation effect of the pump beam that partially depletes the iodine ground state, causing differential absorption and phase delay between the sidebands and the carrier. The beat signal is then mixed down by beating it with the 10.4~MHz local oscillator, resulting in a dispersion signal. 

To further improve the signal-to-noise ratio~(S/N) and remove unwanted offset on the signal, the pump beam is chopped at a frequency of 1~kHz. The derivative signal is then detected using a lock-in amplifier, which selectively amplifies the signal at the chopping frequency, suppressing the noise. The error signal is then fed back directly to a piezo installed on the pre-stabilization cavity steering the frequency on resonance with the iodine transition. The iodine cell was maintained at room temperature (approximately $22^{\circ}$C) and continuously monitored throughout the experiment. At this temperature, we estimate a corresponding iodine partial pressure of about 29.5~Pa. 

To measure the absolute frequency of the stabilized laser, part of the laser radiation (about 1~mW) from the master laser at 1304~nm is transmitted through a 200~m underground fiber cable to a dedicated laboratory equipped with an infrared~(IR) femtosecond fiber frequency comb with 100~MHz repetition rate.
The IR comb light, pre-selected in the spectrum via an optical grating, is superimposed on the 1304~nm optical radiation on a polarization beamsplitter with mutually orthogonal polarizations. The radiation is then sent to a second polarisation beamsplitter, which generates two optical beams detected via a double-balanced fast photodiode. This system allows the production of an RF beat-note signal with a S/N of about 20~dB in a 300~kHz resolution bandwidth. 

While locking the 1304~nm master laser successively to each selected hyperfine $I_2$ transition, the beat-note frequency~\(f_b\) together with comb repetition rate~\(f_{rep}\) and the comb offset frequency~\(f_0\) is measured via dedicated frequency counters. 

Given the low repetition rate of the comb, precise determination of the comb mode, n, for each measurement is done with a standard procedure~\cite{Zhang07} that consists of a pre-measurement with a 0.2~ppm precision wavemeter, followed by at least two sequential measurements of the beat note and the comb parameters, after changing the repetition rate of the comb. The frequency of the iodine transition is then derived by the usual formula: \(f_{I_2}=2\times(n f_{rep} \pm f_b \pm f_0)\), where the two signs are determined in the usual way.


\section{\label{sec:expres}Experimental Results and their Analysis}

\begin{figure}[t]
\centering
\includegraphics[width=\textwidth]{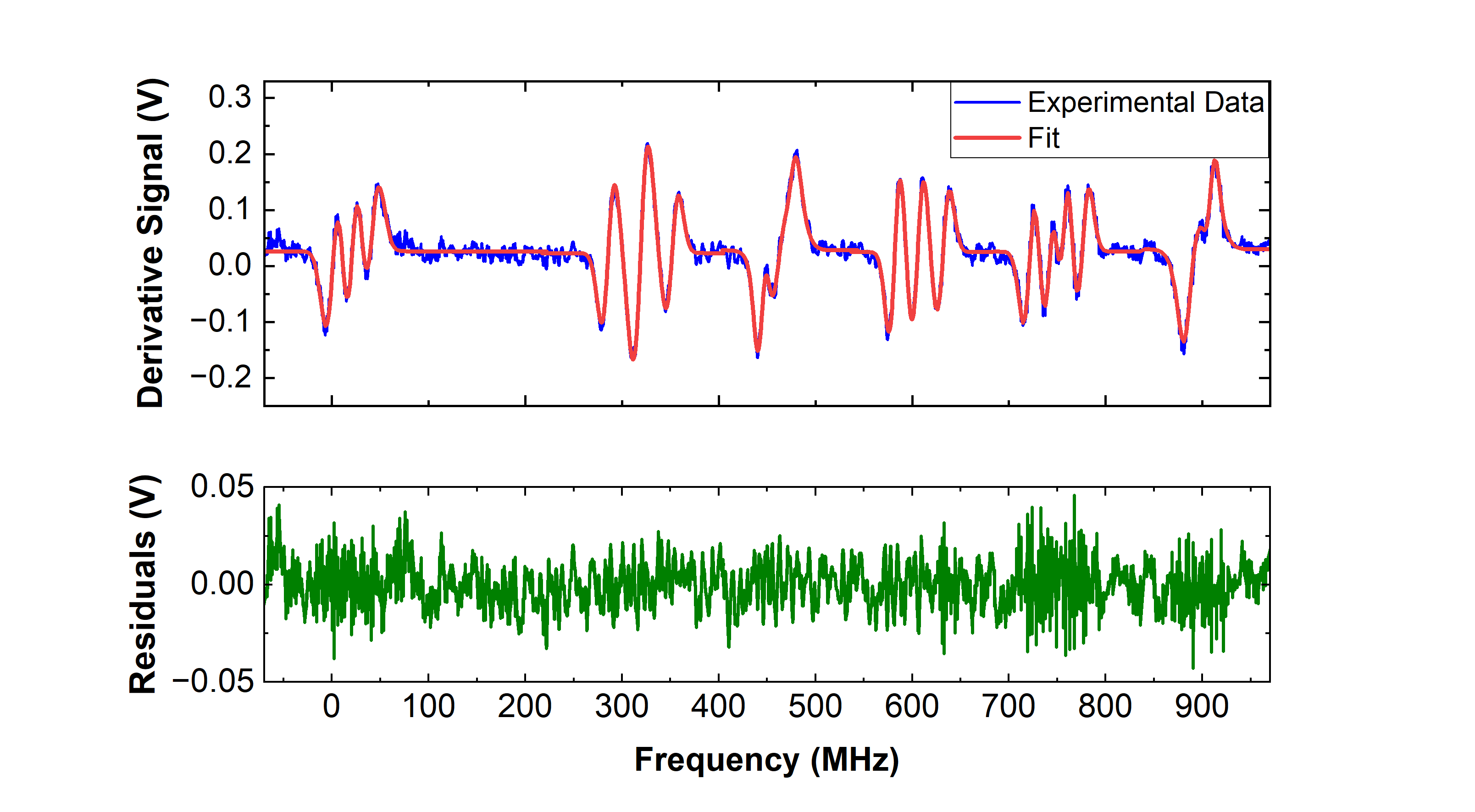}
\caption{(a) Measured derivative signal of the absorption spectra of the hyperfine structures of P(63)~4-4 transition and corresponding fit of the experimental data. (b) Residuals from the fit}
\label{fig:exp_signal}
\end{figure}

Figure~\ref{fig:exp_signal} shows a typical scan of the Doppler-free spectrum of the P(63)~4-4 transition of I$_2$, shown as a blue line. The hyperfine components are labeled as $a_{1}$ - $a_{21}$. Due to the small relative separations between the hyperfine components compared to the linewidth of the observed transitions, well-separated zero crossings for each transition are not always apparent. By using the known absolute frequencies of the hyperfine components, we scaled the frequency of the ramp to estimate the hyperfine splittings. We fit the experimental data (blue line) to a model using a Gaussian derivative for each of the 21 hyperfine components (red line). The model has four free parameters: the Gaussian line width, frequency offset, amplitude, and vertical signal offset.  The relative amplitudes and frequency offsets of the individual components were fixed based on the effective Hamiltonian model. The linewidth obtained from the fit is approximately 7~$\pm$~1~MHz, which corresponds to the observed linewidth under the present experimental conditions. This value likely includes broadening effects from several factors, such as the pump and probe laser intensities, collisional (pressure) broadening, and potential misalignment. While the frequency was stabilized to the zero crossings of these derivative signals, the absolute frequency of the laser was measured with the frequency comb, with an integration time of $60-90$~s. The typical Allan deviation, $\sigma_{y}(\tau)$, ranges from \(2 \times 10^{-11} \)---\(7 \times 10^{-11} \) for different error signals of the hyperfine transitions, measured at an averaging time$(\tau)$ of approximately 20 seconds. All measurements were repeated over a period of one month.

\begin{table*}[htb]
\begin{center}
\captionsetup{justification=justified, singlelinecheck=false}
\caption{Measured absolute frequencies of selected hyperfine components of the P(63)~4-4 transition of the I$_{2}$ molecule at room temperature, obtained in this study. The experimental error values are indicated in parentheses, corresponding to the last digits of the measured frequencies. This table also includes the PGOPHER fit results, as well as literature-derived frequencies using the rovibrational constants and origin values from Ref.~\cite{Luc85}, and hyperfine constants from the empirical formula~\cite{Bodermann2002}. The residuals, which represent the fit error as the difference between the measured and fitted frequencies, are also presented, along with the differences ($\Delta$) between the fitted and literature values. Reduced $\chi^2$ = 0.54 (p-value = 0.78)}
\label{tab:hf_splittings}
\begin{tabularx}{\textwidth}{>{\centering\arraybackslash}p{0.025\linewidth} 
>{\centering\arraybackslash}p{0.025\linewidth} 
>{\centering\arraybackslash}p{0.028\linewidth} 
>{\centering\arraybackslash}p{0.21\linewidth} 
>{\centering\arraybackslash}p{0.17\linewidth} 
>{\centering\arraybackslash}p{0.08\linewidth} 
>{\centering\arraybackslash}p{0.17\linewidth} 
>{\centering\arraybackslash}p{0.06\linewidth}}
\toprule
\small \textbf{HC} & \small \(\boldsymbol{F}'\) & \small \(\boldsymbol{I}\) & \small \textbf{Absolute Frequency (This Work)} & \small \textbf{PGOPHER Fit (This work)} & {\small \textbf{Residuals}} & \small \textbf{Literature values~\cite{Luc85, Bodermann2002}} & {\small \textbf{$\Delta$} } \\
& & & \small \textbf{(MHz)} & \small \textbf{(MHz)} & \small \textbf{(MHz)} & \small \textbf{(MHz)} & \small \textbf{(MHz)} \\

\midrule
a$_1$ & 57 & 5 & 459~523~528.17~(51) & 459~523~528.83 & -0.66 & 459~523~547.75 & 18.92 \\
a$_2$ & 62 & 1 & 459~523~547.48~(35) & 459~523~547.50 & -0.02 & 459~523~566.76 &19.26\\
a$_3$ & 67 & 5 & 459~523~566.3~(13)  & 459~523~565.71 & 0.59 & 459~523~585.31 & 19.61 \\
a$_4$ & 58 & 5 & 459~523~807.23~(22) & 459~523~807.10 & 0.13 & 459~523~826.06 & 18.96 \\
a$_7$ & 66 & 5 & 459~523~874.78~(95) & 459~523~874.14 & 0.64 & 459~523~893.64 & 19.50 \\
a$_{12}$ & 59 & 3 & 459~524~110.43~(19) & 459~524~110.42 & 0.01 & 459~524~129.48 & 19.06 \\
a$_{13}$ & 62 & 3 & 459~524~133.95~(80) & 459~524~134.39 & -0.44 & 459~524~153.64 & 19.25 \\
a$_{14}$ & 65 & 5 & 459~524~160.77~(97) & 459~524~161.13 & -0.36 & 459~524~180.58 & 19.45 \\
a$_{18}$ & 64 & 5 & 459~524~307.77~(34) & 459~524~307.87 & -0.10 & 459~524~327.22 & 19.35 \\
\bottomrule
\end{tabularx}
\end{center}
\end{table*}

\begin{table*}[htb]
\centering
\caption{Hyperfine constants, the origin of the upper vibrational level \((v' = 4)\), and the frequency of the center of gravity ($f_{C.G.}$) for the P(63)~4-4 transition of I$_{2}$. The table reports literature-derived values based on Ref.~\cite{Luc85} and Ref.~\cite{Bodermann2002} alongside the best parameter set obtained from the fit procedure. Error values are indicated in parentheses, corresponding to the last digits of the reported values. Fixed parameters are detailed in the lower section of the table~\cite{Luc85, Bodermann2002}. Double primes (") denote ground state parameters, while the double dagger ($^{\ddag}$) indicates recommended final parameters.}

\vspace*{5mm}
\begin{center}
\begin{tabular}{>{\centering\arraybackslash}m{4cm}>{\centering\arraybackslash}m{3.75cm}>{\centering\arraybackslash}m{3.75cm}>{\centering\arraybackslash}m{7.5cm}}
\toprule
\textbf{Parameters} & \textbf{Fit Results (This work)} & \textbf{Literature values} \\
\midrule
$\Delta eQq$~(MHz) & 1~957.67~(71) & 1~957.672~(25)~\cite{Bodermann2002}$^{\ddag}$ \\
$\Delta C$ ~(kHz) & 20.37(60)$^{\ddag}$ & 21.4~(10)~\cite{Bodermann2002} \\
$f_{(C.G.)}$~(MHz) & 459~524~036.90~(12)$^{\ddag}$ & 459~524~056~(30)~\cite{Luc85} \\
Origin \((v' = 4)\)~(MHz) & 486~021~343.15~(12) & 486~021~362~(30)~\cite{Luc85} \\
\midrule
\textbf{Fixed Parameters} & \multicolumn{2}{c}{\textbf{}} \\
\midrule
\(eQq''\)~(MHz)  & \multicolumn{2}{c}{-2454.46~\cite{Bodermann2002}} \\
\(C''\)~(kHz)  & \multicolumn{2}{c}{3.3~\cite{Bodermann2002}} \\
\(\Delta d\)~(kHz)  & \multicolumn{2}{c}{-6.5~\cite{Bodermann2002}} \\
\(\Delta \delta\)~(kHz)  & \multicolumn{2}{c}{5.9~\cite{Bodermann2002}} \\
\bottomrule
\end{tabular}
\label{tab:HFC}
\end{center}
\end{table*}

In Table~\ref{tab:hf_splittings}, we present the measured absolute frequencies of selected components. The errors indicated in paranthesis represent the statistical uncertainties from independent measurements of each hyperfine component, with a minimum of 4 and a maximum of 10 measurements repeated over a period of one month. The dominant frequency shifts resulting from laser power, vapour pressure, and electronics were determined by changing respectively the relative experimental conditions (laser intensity, iodine cell temperature, and locking offset). All the frequency shifts obtained in these measurements lie within the range of the estimated statistical errors. The table also includes the best-fit results obtained using PGOPHER, along with the corresponding residuals from these fits.  Additionally, the expected values of the hyperfine transition frequencies from literature are provided. These values are calculated using the rovibronic level parameters from Ref.~\cite{Luc85} and the hyperfine constants (\(eQq'\), \(eQq''\), \(C'\) and \(C''\)) from the empirical formulae provided in Ref.~\cite{Bodermann2002}. The discrepancy($\Delta$) between the fitted values and the literature-derived values is shown in the last column. By assigning the individual measured line positions to their corresponding simulated transitions, we performed a fit to extract the hyperfine constants. Table~\ref{tab:HFC} presents the extracted hyperfine constants obtained from this fit. The excited state hyperfine constants \(eQq'\) and \(C'\) and the excited state \(v'= 4\) band origin are fitted to the nine measured transition frequencies. For comparison, we also include the expected values derived from the empirical formula, as well as the band origin values taken from~\cite{Luc85}. In the fit, we fixed the $\Delta d$ and $\Delta\delta$ constants to the values determined from the empirical formula~\cite{Bodermann2002} as $\Delta d = \text{-}6.5~\text{kHz}$ and $\Delta\delta~=~5.9$~kHz. The fitting results suggest that these parameters have a minimal impact on the fit, given the current accuracy of the data. Since the ground state \(v''= 4\) has not been measured previously, the ground state parameters are fixed and only the excited state parameters are treated as free parameters during the fitting process. Specifically, we fix \(eQq''\)~=~-2454.46~MHz and \(C''\)~=~3.3~kHz which is determined using the empirical formula from Ref.~\cite{Bodermann2002}
 
 Although these lower-level constants may not precisely apply to \(v''= 4\), the hyperfine structures' splitting patterns are similar in both ground and excited states, scaled by the differences in the constants \(\Delta eQq (eQq'-eQq'') \), \(\Delta C (C'-C'') \), \(\Delta d (d'-d'') \), and \(\Delta\delta(\delta'-\delta'')\)~\cite{Hanes1971}. Therefore, the differences in the hyperfine constants have a negligible effect on the ground state parameters.  All results in this paper are presented as the differences of the hyperfine constants: $\Delta eQq$, $\Delta C$, $\Delta d$, and $\Delta\delta$. 
 
Additionally, we determined the absolute frequency of the center of gravity ($f_{C.G.}$) of the hyperfine transitions using PGOPHER. This was done by setting the hyperfine constants to zero, with the resulting value also reported in Table~\ref{tab:HFC}. We observed a total shift of around 20~MHz of the center of gravity of the P(63)~4-4 transition with respect to the previously known value~\cite{Luc85}, with a 250-fold reduction of the uncertainty. The PGOPHER code and the accompanying log file summarizing the fitting results, including the fitted parameters (\(eQq\textit{0} \rightarrow eQq\), \(cI \rightarrow C\), \(S \rightarrow -d\), \(J \rightarrow \delta\)), are available in Code 1 (Ref.~\cite{Manzoor2024}). A $\chi^2$-test was performed, resulting in p-values~>~0.78 (significantly above the 5\% threshold). 

To validate our results, we performed additional fitting of the experimental data using SPFIT~\cite{PICKETT91}, an open-source software. The results from SPFIT are consistent with those obtained from PGOPHER, confirming the accuracy of the hyperfine constants derived in our study. Therefore, for accurate prediction of the hyperfine constants investigated in this study, we recommend using the $eQq$ parameter from the empirical formula~\cite{Bodermann2002} and $C$ parameter derived from our experimental values.


\section{Conclusions and Future Prospects} 

We performed precision measurements of the frequency of the hyperfine components of P(63)~4-4 transition of naturally abundant molecular iodine $^{127}$I$_2$ at around 652.4~nm. Based on fitting the experimental data to an effective Hamiltonian model, we deduced the first experimental values of the hyperfine constants $\Delta eQq$ and $\Delta C$. We found reasonable agreement for the $\Delta eQq$ and $\Delta C$ values as compared to the empirical formulae, significantly reducing the uncertainty of $\Delta C$ constant by 40$\%$, thereby improving its accuracy. Using these methods, we also provided the updated absolute frequency of the center of gravity of this transition with an accuracy improved by more than one order of magnitude. Furthermore, we show the second harmonic at 326.8~nm of the radiation stabilized on the hyperfine components of this particular molecular transition is also overlapping with the UV $^1$S$_0$~-~$^3$P$_1$ narrow intercombination transition of cadmium atoms.

This work constitutes an important contribution to the continuous update and improvement of the iodine atlas and the improvement of the precision of the empirical formula, particularly for the spin-rotation constant. 

Building on these results, future work could focus on further refining the empirical formulae by incorporating more precise data from additional iodine transitions that have been studied so far. Additionally, the overlap between the iodine and cadmium transitions offers a valuable opportunity to avoid cumbersome UV setups and improve frequency calibrations for precision spectroscopy on Cd. This transition may also facilitate the development of second-stage cooling of Cd. The use of longer interaction cells or alternative spectroscopic techniques, such as cavity ring-down spectroscopy, could further help resolve hyperfine structures of weak transitions, thereby enhancing the precision and accuracy of future experiments.

\section*{Funding}
This work is partially supported by the European Research Council, (772126-"TICTOCGRAV").  S.M. is supported by project PRIN2022-ISOTOP (Prot. 2022Z8LX9L) funded by MIUR under the "NextGenerationEU" European program. 

\section*{Acknowledgments}
We thank M. Marynowski, and M. Sacco for their participation in the initial stage of the experiment and for participating in the setting up of the 200 m-long fiber link.
S.M. thanks P. \u{C}erm\'ak for fruitful discussions.

\section*{Disclosures}
The authors declare no conflicts of interest.

\section*{Data availability} Data underlying the results presented in this paper are not publicly available at this time but may be obtained from the authors upon reasonable request.

\bibliography{Iodine.bib}

\end{document}